\documentclass[]{mn2e}
\usepackage{times}
\usepackage{epsf}
\usepackage{psfig}
\newif\ifAMStwofonts

\title{The Bicoherence as a Diagnostic for Models of High Frequency QPOs}

\author[Maccarone \& Schnittman] {Thomas J. Maccarone\\ Astronomical
Institute ``Anton Pannekoek'', University of Amsterdam, Kruislaan 403,
1098 SJ, Amsterdam, The Netherlands \newauthor Jeremy D. Schnittman\\Department of Physics, Massachusetts Institute of Technology, 77 Massachusetts Avenue, Cambridge, MA 02139\\}

\date{}

\begin{document}

\maketitle

\label{firstpage}

\def\simlt{\mathrel{\rlap{\lower 3pt\hbox{$\sim$}}
        \raise 2.0pt\hbox{$<$}}}
\def\simgt{\mathrel{\rlap{\lower 3pt\hbox{$\sim$}}
        \raise 2.0pt\hbox{$>$}}}

\input epsf

\begin{abstract}
We discuss the use of the bicoherence - a measure of the phase
coupling of oscillations at different frequencies - as a diagnostic
between different models for high frequency quasi-periodic
oscillations from Galactic black hole candidates.  We show that this
statistic is capable of finding qualitative distinctions between
different hot spot models which produce nearly identical Fourier power density
spectra.  Finally, we show that proposed new timing missions should
detect enough counts to make real use of this statistic.
\end{abstract}


\section{Introduction}
It has long been hoped that X-ray binaries and Active Galactic Nuclei
would prove to be good laboratories for testing general relativity.
The two most promising lines of attack for disentangling relativistic
effects from the physics of accretion disks and of radiative transfer
are high resolution spectroscopy of emission lines (see e.g. Reynolds
\& Nowak 2003 for a review) and studies of high frequency
quasi-periodic oscillations (see e.g. van der Klis 2004 for a review).

Recently, high frequency quasi-periodic oscillations (HFQPOs) have
been found in low mass X-ray binaries for which the black hole masses
are reasonably well measured.  In one case in particular, a frequency
has been identified which is too large to be a Keplerian orbit around
a Schwarzschild black hole (Strohmayer 2001).  Furthermore, a pattern
has begun emerging where these high frequency QPOs are found in pairs
with 2:3 ratios of frequencies.  This was first pointed out by
Abramowicz \& Kluzniak (2001), based on combining observational
results from Strohmayer (2001) and Remillard et al. (1999). More
recent work from Miller et al. (2001), Remillard et al. (2003),
Remillard et al. (2004) and Homan et al. (2004) has lent more weight
to the idea that 2:3 frequency ratios are quite common in these
systems.

A variety of theoretical models have been developed to explain these
QPOs, most requiring a spinning black hole, but often requiring rather
different values of the spin (compare, e.g. Abramowicz \& Kluzniak
2001; Rezzolla et al. 2003; Li \& Narayan 2004).  Therefore, there is
still much ``astrophysics'' (i.e. physics of disk structure and
stability and physics of radiative transfer) that must be understood
before the fundamental physics can be probed in these systems, but
there is strong cause for optimism that these systems really will
ultimately tell us something profound about spinning black holes.

A key first step to disentangling the ``astrophysics'' is, of course,
to develop models which not only match the important frequencies,
but also include radiation mechanisms such that the observed X-rays would
actually be modulated at that frequency.  Such has been done recently
for a particular realization of the case of the parametric resonance
model of Abramowicz \& Kluzniak (2001), by considering the possibility
of hot spots that form at the resonant radii in the accretion disk
(Schnittman \& Bertschinger 2004; Schnittman 2004).  These authors
have found that different sets of model parameters can produce the same
Fourier power density spectrum with dramatically different qualitative
appearances to the light curves.  In this Letter, we will show that
higher order variability statistics, particularly the bispectrum, can
break this degeneracy.

\section{Statistical Methods}

One key way to distinguish between different mechanisms which produce
the same power spectra from qualitatively different light curves is to
study the non-linearity of the variability.  Linear variability is
that in which the phases at the different frequencies in the Fourier
spectrum of a time series are uncorrelated with one another, while
time series with non-linear variability show Fourier spectra with
correlations between the phases at different frequencies.  Some
particularly useful tools for studying non-linearity are the
bispectrum and the closely related bicoherence.

The bispectrum computed from a time series broken into $N$ segments is
defined as:
\begin{equation}
B(k,l)=\frac{1}{N} \sum_{i=0}^{N-1} X_i(k)X_i(l)X^*_i(k+l),
\end{equation}
where $X_i(k)$ is the frequency $k$ component of the discrete Fourier
transform of the $i$-th time series (e.g. Mendel 1991; Fackrell 1996
and references within).  It is a complex quantity that measures the
strength of the phase coupling of different Fourier frequencies in a
light curve and has a phase of its own which is the sum of the phases
at the two lower frequencies minus the phase at the highest frequency.
Its value is unaffected by additive Gaussian noise, although its
variance will increase for a noisy signal.

A related quantity, the bicoherence is the vector magnitude of the
bispectrum, normalised to lie between 0 and 1.  Defined analogously to
the cross-coherence function (e.g. Nowak \& Vaughan 1996), it is the
vector sum of a series of bispectrum measurements divided by the sum
of the magnitudes of the individual measurements.  If the biphase (the
phase of the bispectrum) remains constant over time, then the
bicoherence will have a value of unity, while if the phase is random,
then the bicoherence will approach zero in the limit of an infinite
number of measurements.   Mathematically, the bicoherence $b$ is defined as:
\begin{equation}
b^2(k,l) = \frac{\left|\sum_{i=0}^{N-1}{X_i(k)X_i(l)X^*_i(k+l)}\right|^2}{\sum_{i=0}^{N-1}{\left|X_i(k)X_i(l)\right|^2}\sum_{i=0}^{N-1}{\left|X_i(k+l)\right|^2}}.
\end{equation}
This quantity's value is affected by Gaussian noise, but it can be
considerably more useful than the bispectrum itself for determining
whether two signals are coupled non-linearly.  In an
astronomical time series analysis context, it has been previously
applied to the broad components in the power spectra of Cygnus~X-1 and
GX~339-4, in both cases finding non-linear variability through the
presence of non-zero bicoherences over a wide range of frequencies
(Maccarone \& Coppi 2002).

Since that work, we have become aware of a correction which is, in
principle, important for studying aperiodic variability with the
bicoherence, namely that the maximum value of the bicoherence is
suppressed by smearing of many frequencies into a single bin in the
discrete Fourier transform.  This suppression cannot be calculated in
a straightforward way (see e.g. Greb \& Rusbridge 1988).  However, we
also note that since comparisons in Maccarone \& Coppi (2002) were
made only with model calculations made with the same time binning as
the real data, these effects, whatever they may be, are the same for
the real data and the simulated data, and hence the conclusions of
that paper are not affected substantially.

\section{A Brief Review of the QPO Models}

There exist, at the time of this paper's writing, at least four basic
concepts for producing the high frequency quasi-periodic oscillations
seen from accreting black holes.  In historical order, these are
diskoseismology (e.g. Okazaki, Kato \& Fukue 1987), relativistic
coordinate frequencies (e.g. Stella \& Vietri 1999), Rayleigh-Taylor
instabilities (e.g. Titarchuk 2002, 2003; Li \& Narayan 2004) and
oscillating tori (Rezzolla et al. 2003).  We will briefly summarize
the properties of the models in a different order in an effort to
smooth the flow of the paper.

The first model proposed was based on diskoseismology - the excitation
of various trapped modes in the inner region of a Keplerian accretion
disk in a relativistic potential (e.g. Okazaki, Kato \& Fukue 1987;
Nowak et al. 1997).  This model seems not to be directly applicable to
the data, at least for the cases where small integer ratios of
frequencies exist; it would require considerable fine tuning in the
different mass and spin values for the black holes to produce
routinely a 2:3 frequency ratio.  Chen \& Taam (1995) showed that a
slim disk (see e.g. Abramowicz et al. 1988) can produce oscillations
at a frequency very close to the radial epicyclic frequency of the
accretion disk, sometimes with power at twice this frequency; it also
appears that there may be some excess power at 3/2 of the epicyclic
frequency in one of the simulations presented in Chen \& Taam (1995)
-- see Figure 5 of that paper -- but it is likely that a longer
hydrodynamic simulation would be needed to confirm this.  In a similar
vein is the oscillating torus model of Rezzolla et al. (2003).  This
model also applies calculations of the frequencies of $p$-modes
(i.e. sound waves) of Zanotti, Rezzolla \& Font (2003), but in a
geometrically thick, pressure supported torus (as expected at high
accretion rates like those where the HFQPOs are seen - De Villiers,
Krolik \& Hawley 2003; Kato 2004), rather than in a geometrically
thin, Keplerian accretion disk.  In this case, the different overtones
are found to be approximately in a series of integer ratios, starting
from 2, so the model is compatible with existing data on high
frequency QPOs in black holes.

The model most recently applied to high frequency QPOs from black hole
candidates is that of Rayleigh-Taylor instabilities, although the same
basic idea had previously been applied to QPOs from accreting neutron
stars (Titarchuk 2002, 2003).  In this picture, non-axisymmetric
structures can grow unstably at the magnetospheric radius (presumed to
exist also for black holes, as their accretion disks can become
magnetically dominated) with frequencies of integer ratios of the
angular frequency at that radius, though the lowest mode will be
stable for low gas pressures (Li \& Narayan 2004).

After the first indications that small integer ratios between HFQPO
frequencies were likely, it was noted by Abramowicz \& Kluzniak (2001)
that if the relativistic coordinate frequencies determined the
frequencies of the quasi-periodic oscillations (see e.g. Stella \&
Vietri 1999) then resonances between these different frequencies
(e.g. vertical and radial epicyclic frequencies) might occur at
locations in the accretion disk where these frequencies have small
integer ratios.  More recently, Schnittman \& Bertschinger (2004)
performed ray tracing calculations of the light curve of sheared hot
spots produced with a 1:3 radial to azimuthal epicyclic frequency
ratio and found good agreement with the observed locations of the
power spectrum peaks and their relative amplitudes, while Bursa et
al. (2004) have shown that it is also possible to produce these
frequencies from radial and vertical oscillations of a torus located
such that the radial epicyclic frequency is 2/3 the orbital frequency.
Schnittman (2004) extended the work by considering the effects of
multiple hot spots under different conditions in order to broaden the
\textit{periodic} oscillations computed in Schnittman \& Bertschinger
(2004) into the \textit{quasi}-periodic oscillations that are actually
observed.  In this paper, we present bicoherence calculations only for
this last model for the simple reason that this is the only model for
which simulated light curves are currently available.  As simulated
light curves for other models become available, we will consider their
higher order statistical properties as well.

\section{The Bicoherence of the Simulated Data}
We now apply the bicoherence to the simulated data.  We consider two
different model calculations from Schnittman (2004) which give nearly
identical power spectra. In each case, the quasi-periodic oscillations
are produced by a 1:3 resonance between the radial epicyclic frequency
and the orbital frequency.  The parameters have been chosen such that
the orbital frequency is 285 Hz, and the radial epicyclic frequency is
95 Hz.  This corresponds to a black hole mass of 10 $M_\odot$ and a
spin $a/M=0.5$, with the resonance occuring at a radius of 4.89$M$ in
geometrized units where $G=c=1$; the black hole mass and QPO
frequencies compare reasonably well to those observed for XTE
J~1550-564 (see e.g. Miller et al. 2001; Remillard et al. 2002).  The
disk inclination is also fixed to be 70 degrees (where 90 degrees is
an edge-on disk); this parameter does not affect the frequencies
observed, but can affect the amplitudes of the QPOs in the context of
the model we are considering here (Schnittman \& Bertschinger 2004).
In each case we compute 1000 seconds of simulated data with a binning
timescale of the lightcurve of 0.1 msec.  We then compute Fourier
transforms by breaking the data into 2440 segments of 4096 data points,
making use of 999.424 seconds of the simulated data.

\begin{figure*}
\centerline{\psfig{figure=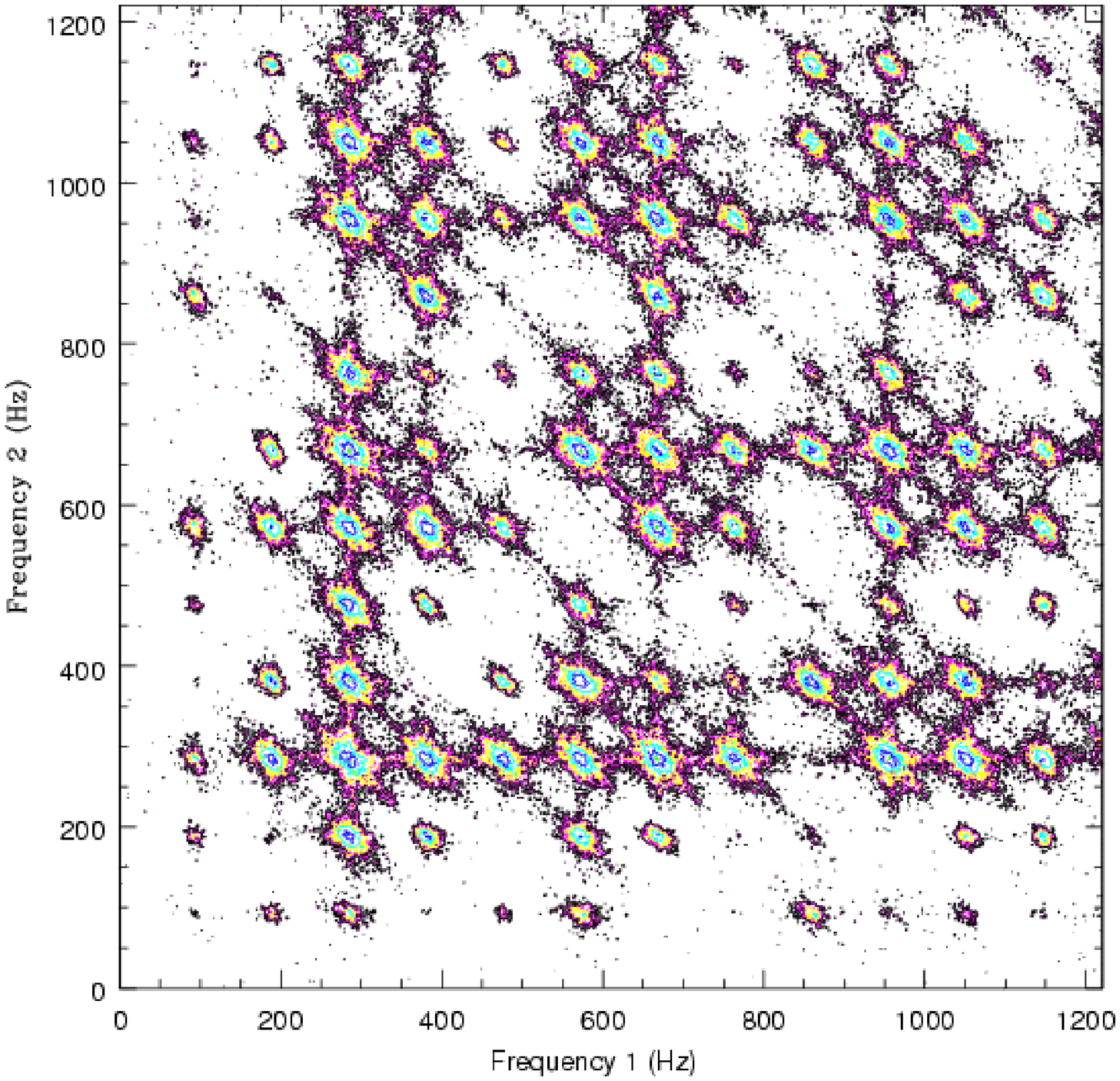,height=8 cm}\psfig{figure=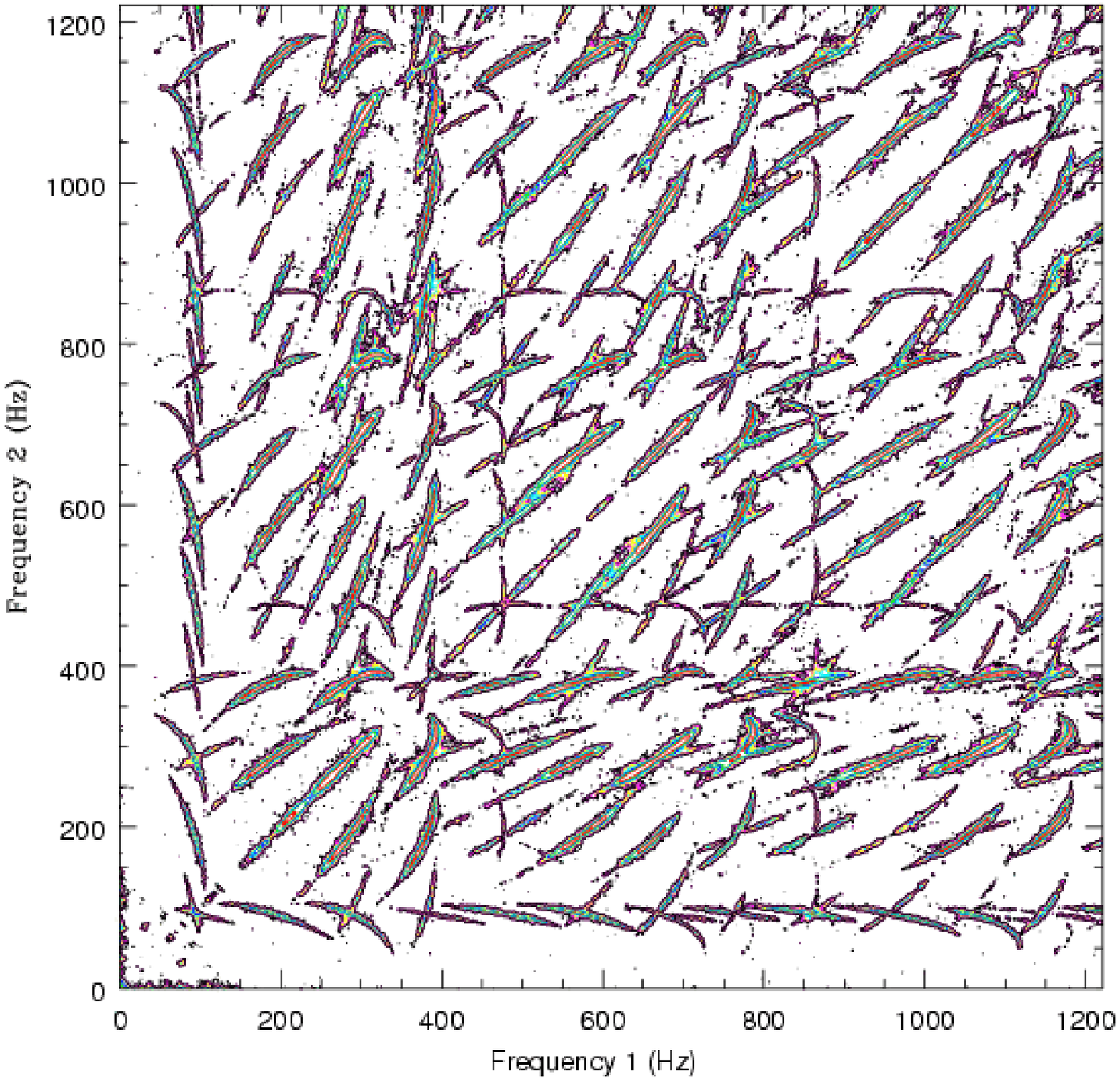,height=8cm}}
\caption{The bicoherences for case 1 (left) and case 2 (right), with
no Poisson statistics corrections made. The contour levels for the
squared bicoherence, $b^2$ are
$10^{-1.0},10^{-1.5},10^{-1.75},10^{-2.0},10^{-2.25},$ and
$10^{-2.65},$ in the colors red, green, dark blue, light blue, yellow,
and purple, respectively.  The frequencies correspond to a $10M_\odot$
black hole with spin parameter $a/M=0.5$.  Note that the symmetry
through the line $x$=$y$ is trivial.}
\label{sim}
\end{figure*}

In the first case, short lived hot spots exist with their orbits all
at a single radius, being continually created and destroyed with a
characteristic lifetime of 4 orbits.  In the second case, long-lived
(lifetimes of 100 msec, or about 30 orbits) hot spots are distributed
over a range of radii ($\delta r \approx 0.05M$).  In both cases, the
hot spots are on orbits with eccentricities of 0.1.  For each model,
the variability appears quasi-periodic, rather than truly periodic,
but for different reasons.  In the first case, the creation and
destruction of hot spots on short timescales leads to a phase jitter
in the light curves.  These random offsets in phase broaden the
observed periodicity.  In the second case, the power spectrum is truly
showing that there are many periodicities in the system, with coherent
phases.  The bicoherence easily detects this difference, as can be
seen from Figure \ref{sim}.  In case 1, the bicoherence shows
``circular'' peaks at various combinations of frequencies where there
is power at $f_1$, $f_2$ and $f_1+f_2$ in the contour plot,
essentially delta function peaks convolved with two-dimensional
Lorenzians due to the random phase broadening.  In case 2, the
bicoherence shows thin elongated peaks, oriented in a variety of
directions depending on $f_1$ and $f_2$.

The reason for this difference is straightforward.  In the first case,
all hot spots have the same geodesic frequencies, so during a hot
spot's lifetime, it is phase locked to all the other hot spots, giving
a collection of delta function peaks at the coordinate
frequencies. The random phase jitter will broaden the
$\delta-$functions into QPOs, with a similar Lorentzian width as
described in Schnittman (2004).  The hot spots being created and
destroyed in the middle of a Fourier transform window will thus create
leakage in the power of the QPO to frequencies near the central
frequency, but there will be a phase relation between the power in
these frequencies and the phase in the central frequency.  The phase
jitter should thus provide a broadening in the bicoherence similar to
that in the power spectrum.  We note that the peaks do appear to be
somewhat elongated, with the direction of elongation such that the
sums of the two smaller frequencies equal the centroid frequencies of
the highest frequency QPO in the triplet.  This is likely because the
centroid is the only truly physical frequency in this case, but a
rigorous proof of this point is beyond the scope of this paper.

\begin{figure*}
\centerline{\psfig{figure=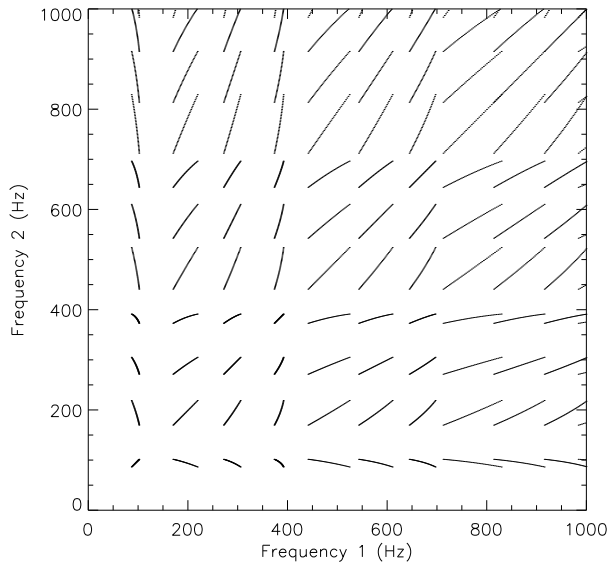,height=8 cm}}
\caption{The tracks showing how different harmonics of the QPO vary
with respect to one another when the radius of the hot spot orbit
varies around a central value $r\approx r_0\pm0.2M$.  The
frequencies correspond to a $10M_\odot$ black hole with spin parameter
$a/M=0.5$.  Note that the symmetry through the line $x$=$y$ is trivial.}
\label{analytical}
\end{figure*}

In the second case, where there are many frequencies in the power
spectrum due to hot spots found over a range of radii, there will be
phase coherence between the different harmonics of each individual hot
spot, but not with the hot spots at slightly different frequencies.
There will thus be bicoherence between the various harmonic
frequencies found at any individual radius, but not between
frequencies found at different radii.  This second case could be
especially interesting.  We have calculated analytically the ratios
expected between different harmonics' frequencies if the radius at
which the hot spot occurs is allowed to vary, and have plotted them in
Figure \ref{analytical}.  If in real data, similar tracks are seen,
then, in the context of this model, they would give the relationships
between the different relativistic frequencies.  Since these tracks
trace the coordinate frequencies as a function of radial distance from
the black hole, they may be used to make precise measurements of the
black hole's mass and spin, plus the central radius of the
perturbations.  Since the observed QPOs are rather narrow, so this
method would probably be of use only with high signal-to-noise data.

\begin{figure*}
\centerline{\psfig{figure=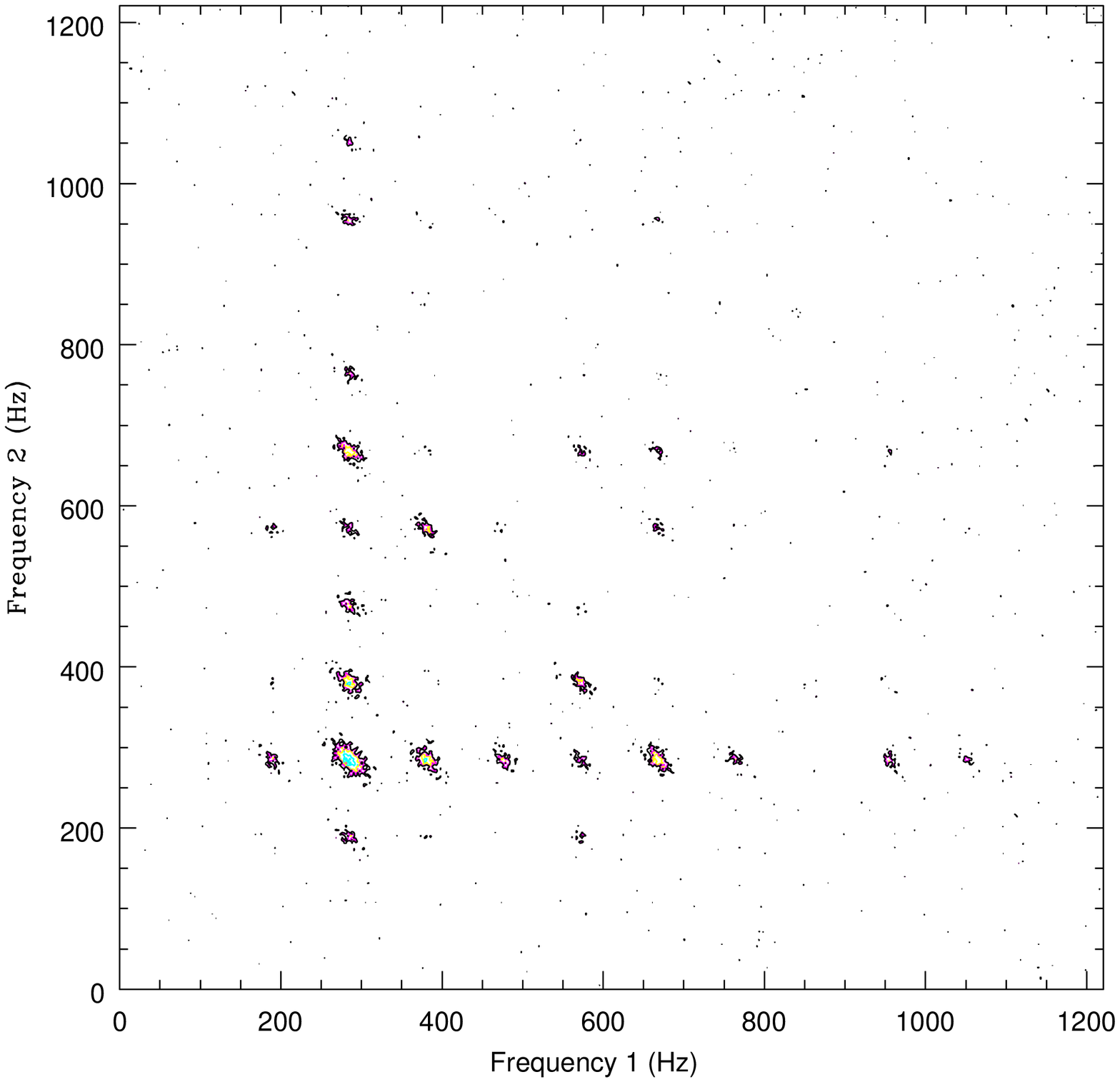,height=8cm} \psfig{figure=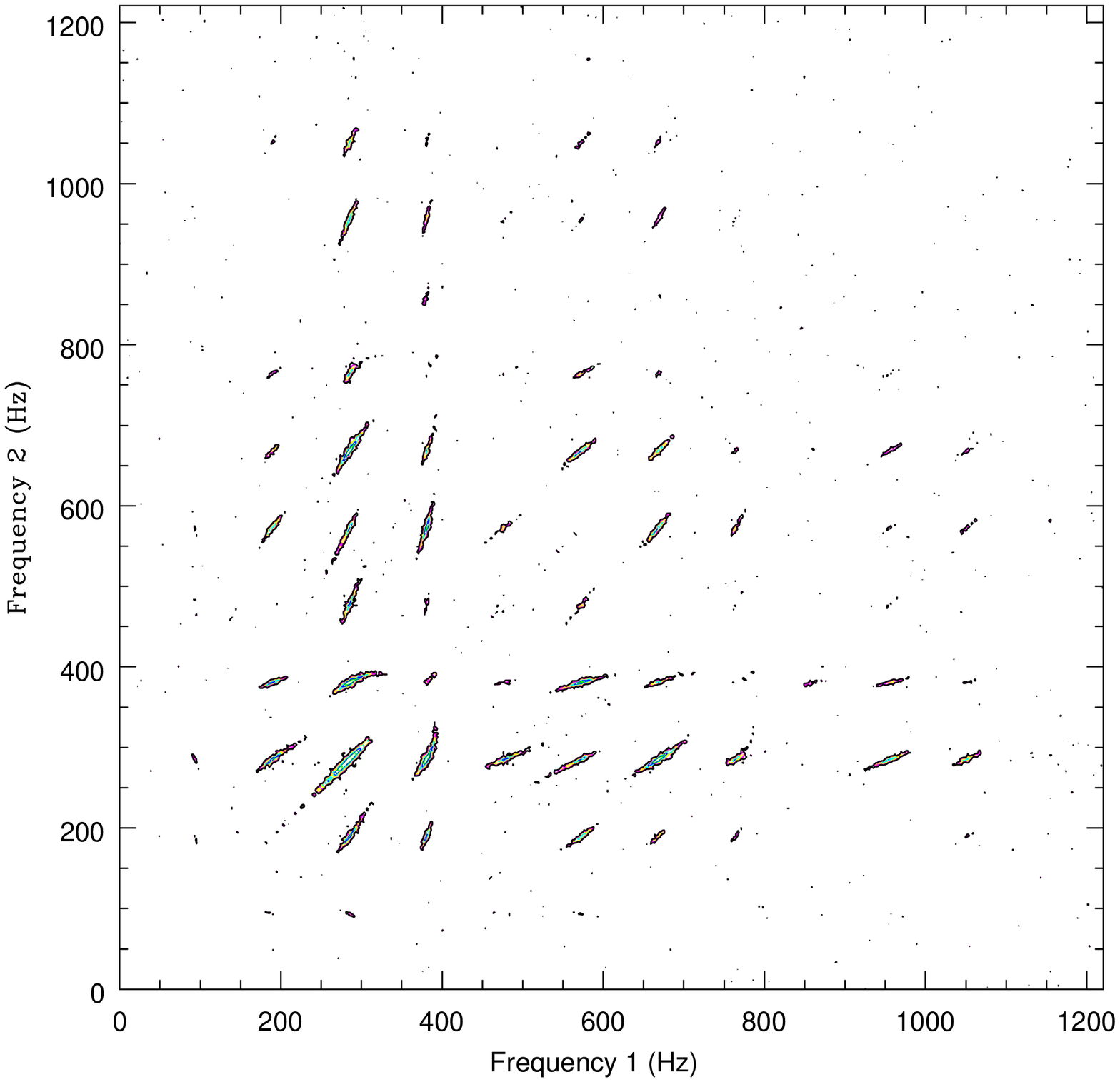,height=8cm}}
\caption{The bicoherences for case 1 (left) and case 2 (right),
assuming 30 m$^2$ area for the detector and a 1000 second integration.
The contour levels for the squared bicoherence, $b^2$ are
$10^{-1.0},10^{-1.5},10^{-1.75},10^{-2.00},10^{-2.25},$ and
$10^{-2.65},$ in the colors red, green, dark blue, light blue, yellow,
purple, and black, respectively. The frequencies correspond to a
$10M_\odot$ black hole with spin parameter $a/M=0.5$.  Note that the
symmetry through the line $x$=$y$ is trivial.}
\label{xeus}
\end{figure*}
\section{Simulations with Poisson Noise}

To consider whether this observational test is really feasible, we
have performed simulations with the rms amplitude of the oscillations
reduced to realistic levels and with Poisson noise added.  We consider
two count rate regimes - one similar to that detected by {\it RXTE}
for the typical X-ray transients at about 10 kpc, which is about
10,000 counts per second, and another which would be expected from the
same source, but with a 30 m$^2$ detector.  In each case, we allow 6\%
of the counts to come from the variable component and to have,
intrinsically, count rates given by the simulated light curves of
Schnittman (2004), and the remaining 94\% of the counts to come from a
constant component.  We then simulate observed numbers of counts in
100 microsecond segments as Poisson deviates (Press et al. 1992) of
the model count rates.  

We then compute the bicoherence as above, but with 2440 segments of
4096 data points, for a total of 999.424 seconds of simulated data.
For the {\it RXTE} count rates, we find that the bicoherence plots
show only noise and only the strongest peak in the power spectrum is
clearly significant in a 1000 second simulated observation, while
marginal detections exist for the QPOs at two-thirds of and twice this
frequency ($\nu_\phi-\nu_r$ and $2\nu_\phi$).  This is as expected
based on real data, which generally requires exposure times much
longer than 1000 seconds to detect these QPOs.  Longer simulated light
curves have not been computed at this time due to computational
constraints.  However, since the signal-to-noise in the bicoherence
should be substantially worse than the signal-to-noise in the power
spectrum, bicoherence measurements should be possible only when a peak
in the power spectrum is considerably stronger than the Poisson level.

For the count rates expected from a 30 m$^2$ detector, we find that
even within 1000 seconds, several of the higher (i.e. $n>4$) harmonics
are observable in the power spectrum and show the clear elongation in
the bicoherence plot for case 2, indicating that proposed missions
should be capable of making use of the bicoherence for studying
HFQPOs.  A few very weak peaks are seen in the bicoherence in case 1
even in 1000 seconds.  The simulated bicoherences for a 30 $m^2$
detector are plotted in Figure \ref{xeus}.  We note that these simulations are
a bit over-simplified, in that we have not included the lower
frequency QPOs and low frequency band-limited noise that are typically
observed in conjunction with the HFQPOs, but that these variability
components should not significantly affect the 
phase coupling of the high frequency QPOs.  We also note that it might
be possible to make use of the bicoherence even with {\it RXTE} if a
more nearby X-ray transient goes into outburst, but that in such a
case, the deadtime effects we have neglected here might become
important.

\section{Conclusions}

We have shown here that the bicoherence can be useful in breaking the
degeneracies between different QPO models which produce the same
Fourier power spectrum.  In particular, we have shown that in the
context of a resonance model for the high frequency quasi-periodic
oscillations seen from accreting black holes, the bispectrum is
capable of distinguishing between broadening due to phase jitter
caused by the rapid creation and destruction of hot spots and
broadening due to an intrinsic distribution of physical frequencies in
a broad range around a central value.  In future work, we will examine
the properties of the bicoherence for other models for these QPOs,
such as oscillations in a pressure supported torus (Rezzolla et
al. 2003).  We note though that this diagnostic is likely to be useful
only with new instrumentation (or the fortuitous outburst of a bright
X-ray transient within about 3 kpc of the Sun).  At the typical 10~kpc
distances of X-ray transients, $RXTE$ is capable of detecting these
QPOs generally only with rather long intergrations and careful
selection of the photon energy bands to maximize the signal-to-noise,
but proposed timing missions with considerably larger collections
areas, like XTRA (e.g. Str\"uder et al. 2004) and the Relativistic
Astrophysics Explorer (e.g. Kaaret 2002) should have considerably
greater potential for making use of these statistics.

\section{Acknowledgments}  
TM wishes to thank Marek Abramowicz, Wlodek Kluzniak, Shin Yoshida and
Olindo Zanotti for stimulating discussions re-motivating his interest
in X-ray variability; Phil Uttley and Simon Vaughan for discussions of
statistical properties of the bispectrum; Mariano Mendez and Marc
Klein Wolt for discussions of observational properties of HFQPOs; Ron
Elsner for providing some useful background information on the
bispectrum; and Luciano Rezzolla and Michiel van der Klis for comments
on the manuscript as well as additional useful discussions.  JDS would
like to thank Edmund Bertschinger and Ron Remillard for many useful
discussions, and would like to acknowledge support from NASA grant
NAG5-13306.

\label{lastpage}
\end{document}